%% file: kormendy.tex
\documentstyle[epsfig]{aipproc}

\begin{document}
\null\vskip -40pt
\title{Supermassive Black Holes\\ in Galactic Nuclei}

\author{John Kormendy$^*$ and Karl Gebhardt$^*$}
\address{$^*$Department of Astronomy, RLM 15.308, University of Texas,
      Austin, TX 78712\\
\vskip 5pt
To appear in The 20th Texas Symposium on Relativistic Astrophysics,
ed.~H.~Martel \& J.~C.~Wheeler, AIP, in press}

\maketitle

\vskip -1pt
\begin{abstract}

We review the motivation and search for supermassive black holes (BHs) in
galaxies.  Energetic nuclear activity provides indirect but compelling evidence
for BH engines. Ground-based dynamical searches for central dark objects are 
reviewed in Kormendy \& Richstone (1995, ARA\&A, 33, 581).  Here we
provide an update of results from the {\it Hubble Space Telescope\/} (HST).
This has greatly accelerated the detection rate.  As of 2001 March, dynamical 
BH detections are available for at least 37 galaxies.  

The demographics of these objects lead to the following conclusions: 
(1) BH mass correlates with the luminosity of the bulge component of the
    host galaxy, albeit with considerable scatter.  The median BH mass fraction
    is 0.13\thinspace\% of the mass of the bulge.  
(2) BH mass correlates with the mean velocity dispersion of the bulge 
    inside its effective radius, i.{\thinspace}e., with how strongly
    the bulge stars are gravitationally bound to each other.  For the best 
    mass determinations, the scatter is consistent with the measurement errors.
(3) BH mass correlates with the luminosity of the high-density central
    component in disk galaxies independent of whether this is a real bulge 
    (a mini-elliptical, believed to form via a merger-induced dissipative
    collapse and starburst) or a ``pseudobulge'' (believed to form by inward
    transport of disk material). 
(4) BH mass does not correlate with the luminosity of galaxy disks.  If pure
    disks contain BHs (and active nuclei imply that some do), then their masses
    are much smaller than 0.13\thinspace\% of the mass of the disk.

We conclude that present observations show no dependence of BH mass on the
 details of whether
    BH feeding happens rapidly during a collapse or slowly via secular evolution
    of the disk.
The above results increasingly support the hypothesis that the major events
    that form a bulge or elliptical galaxy and the main growth phases of its
    BH -- when it shone like a quasar -- were the same events. 
\end{abstract}

\newdimen\sa  \def\sd{\sa=.1em  \ifmmode $\rlap{.}$''$\kern -\sa$
                                \else \rlap{.}$''$\kern -\sa\fi}
\def\ts{\thinspace}
\def\etal{{\it et~al.~}} \def\etnuk{{\it et~nuk.~}}
\def\gapprox{$_>\atop{^\sim}$}
\def\lapprox{$_<\atop{^\sim}$}

\section*{Motivation}

      Black holes (BHs) progressed from a theoretical concept to a necessary
ingredient in extragalactic astronomy with the discovery of quasars by Schmidt
(1963). Radio astronomy was a growth industry at the time; many radio sources
were identified with well-known phenomena such as supernova explosions.  But a 
few were identified only with ``stars'' whose optical spectra showed nothing
more than broad emission lines at unfamiliar wavelengths.  Schmidt discovered
that one of these ``quasi-stellar radio sources'' or ``quasars'', 3C 273, had a
redshift of 16\ts\% of the speed of light.  This was astonishing: the Hubble
law of the expansion of the Universe implied that 3C 273 was one of the most
distant objects known.  But it was not faint.  This meant that 3C 273 had to be
enormously luminous -- more luminous than any galaxy.  Larger quasar redshifts
soon followed.  Explaining their energy output became the first strong argument
for gravity power (Zel'dovich 1964; Salpeter 1964).

      Studies of radio jets sharpened the argument. Many quasars and lower-power
active galactic nuclei (AGNs) emit jets of elementary particles that are
prominent in the radio and sometimes visible at optical wavelengths.  Many are
bisymmetric and feed lobes of emission at their ends (e.{\ts}g., Fig.~1).  Based
on these, \hbox{Lynden-Bell} (1969, 1978) provided a convincing argument for
gravity power.  Suppose that we try to explain the typical quasar using nuclear
fusion reactions, the most efficient power source that was commonly studied at
the time.  The total energy output of a quasar is at least the energy stored
in its radio halo, $E \sim 10^{54}$ J.  Via $E = mc^2$, this energy weighs
$10^7$ solar masses ($M_\odot$).  But nuclear reactions have an efficiency of
only 0.7\ts\%.  So the mass that was processed by the quasar in order to convert
$10^7$ $M_\odot$ into energy must have been $10^9$ $M_\odot$.  This waste mass
became part of the quasar engine.  Meanwhile, rapid brightness variations showed
that quasars are tiny, with diameters $2R \lesssim 10^{13}$ m.  But the
gravitational potential energy of $10^9$ $M_\odot$ compressed inside $10^{13}$ m
is $GM^2/R \sim 10^{55}$ J.  ``Evidently, although our aim was to produce a
model based on nuclear fuel, we have ended up with a model which has produced
more than enough energy by gravitational contraction.  The nuclear fuel has
ended as an irrelevance'' (Lynden-Bell 1978).  This argument convinced many
people that BHs are the most plausible quasar engines.

\vfill

\includegraphics{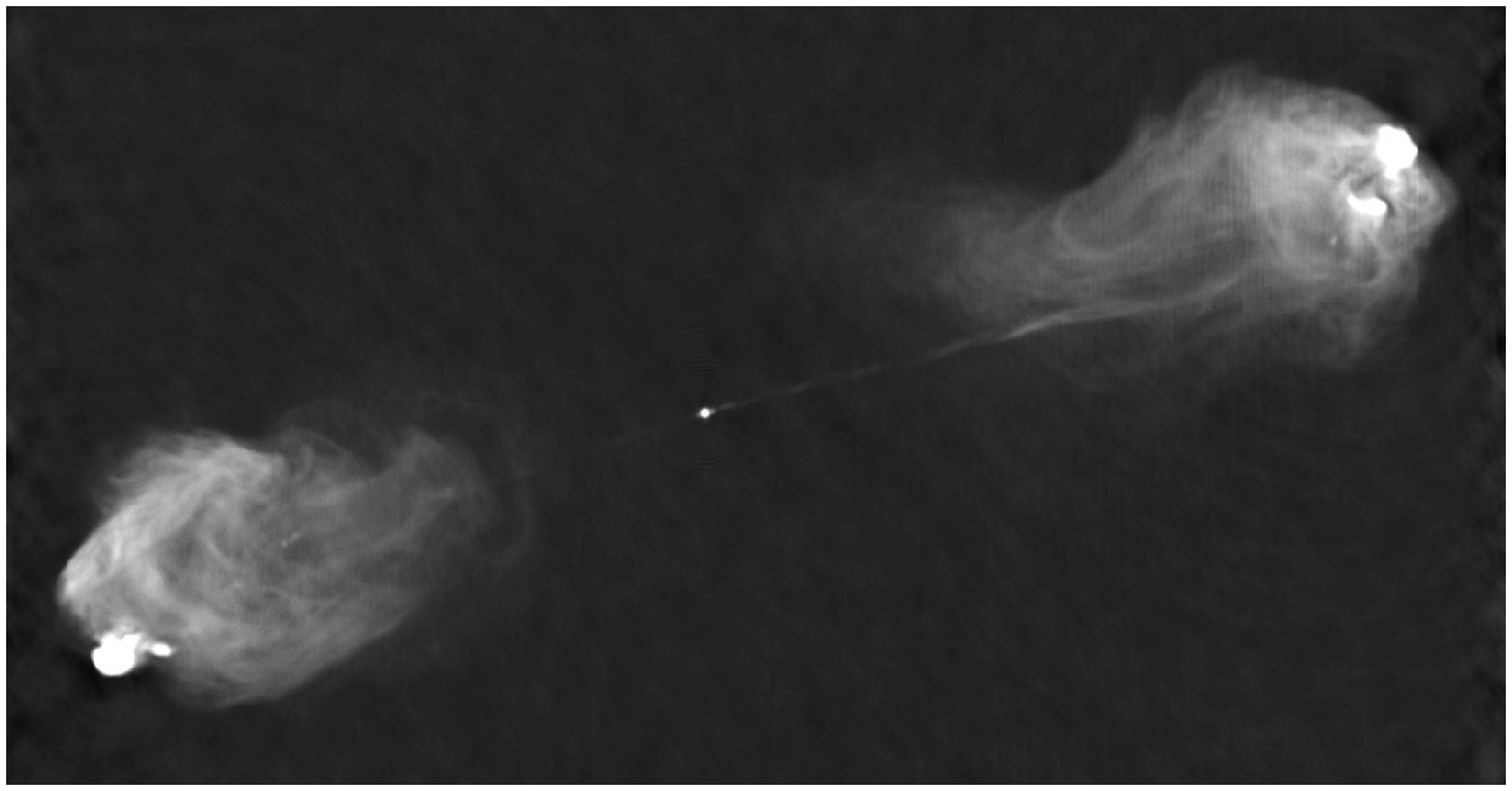}

{\bf FIGURE 1.} Cygnus A at 6 cm wavelength (Perley, Dreher, \& Cowan 1984).
The central point source is the galaxy nucleus; it feeds oppositely directed
jets (only one of which is easily visible at the present contrast) and lobes
of radio-emitting plasma.  The resolution of this image is about 0\sd4.

\eject

      Jets also provide more qualitative arguments.  Many are straight over
$\sim$\ts$10^6$~pc in length.  This argues against the most plausible
alternative explanation for AGNs, namely bursts of supernova explosions.  The
fact that jet engines remember ejection directions for $\gtrsim$\ts$10^6$ yr is
suggestive of gyroscopes such as rotating BHs.  Finally, in many AGNs, jet knots
are observed to move away from the center of the galaxy at apparent velocities
of several times the speed of light, $c$.  These can be understood if the jets
are pointed almost at us and if the true velocities are almost as large as $c$
(Blandford, McKee, \& Rees 1977).  Observations of superluminal motions provide
the cleanest argument for relativistically deep potential wells.

      By the early 1980s, this evidence had resulted in a well-established
paradigm in which AGNs are powered by BHs accreting gas and stars (Rees 1984).
Wound-up magnetic fields are thought to eject particles in jets along the
rotation poles.  Energy arguments imply masses $M_\bullet \sim 10^6$ to
$10^{9.5}$ $M_\odot$, so we refer to these as supermassive BHs to distinguish
them from ordinary-mass (several-$M_\odot$) BHs produced by the deaths of
high-mass stars. But despite the popularity of the paradigm, there was no direct
dynamical evidence for supermassive BHs.  The black hole search therefore became
a very hot subject.  It was also dangerous, because it is easy to believe that
we have proved what we expect to find.  Standards of proof had to be very high.

\section*{The Search for Supermassive Black Holes}

    Kormendy \& Richstone (1995) review BH search techniques and summarize the
ground-based detections.  Recent reviews (e.{\ts}g., Richstone \etnuk 1998)
concentrate on BH astrophysics.  There has not been a comprehensive review of BH
discoveries made with the {\it Hubble Space Telescope\/} (HST), so we provide a 
summary here.

      There are ground-based BH detections in 10 galaxies, including
the nearest (our Galaxy, M{\ts}31, and M{\ts}32) and the best (our Galaxy and
NGC 4258) candidates.  References are given in Kormendy \& Richstone (1995)
and Table 1.  Of the 7 stellar-dynamical cases, 6 have been
reobserved with HST.  In all cases, the BH detection was confirmed and the
ground-based BH mass agrees with the HST result to within a factor of
$\sim$\ts2.  The HST papers are:
M{\ts}31: Statler \etal (1999), Bacon \etal (2001); M{\ts}32: van der Marel
\etal (1998); NGC 3115: Kormendy \etnuk (1996a), Emsellem \etal (1999); 
NGC 3377: Richstone \etnuk (2001), NGC 4486B (Green \etal 2001), and NGC 4594: Kormendy \etnuk (1996b).  

      Our Galaxy is the strongest BH case, based on observations of velocities
in the plane of the sky of stars in a cluster within $0\sd5$ = 0.02 pc of the
central \hbox{radio source} Sgr A* (Eckart \& Genzel 1997; Genzel \etal 1997,
2000; Ghez \etal 1998, 2000).  The fastest star is moving at $1350 \pm 40$ km
s$^{-1}$.  Acceleration vectors have been measured for three stars; they
intersect, to within the still-large errors, at Sgr A*, supporting the 
identification of the radio source with the inferred central mass of $M_\bullet
= (2.6 \pm 0.2) \times 10^6$ $M_\odot$ (Ghez \etal 2000).  The stellar orbital
periods could be as short as several decades, so we can look forward
to seeing the Galactic center rotate in our lifetimes!  Most important, the mass 
$M_\bullet$ is constrained to live inside such a small radius that alternatives
to a supermassive black hole are ruled out by astrophysical constraints.  Brown
dwarf stars would collide, merge, and become luminous, and clusters of white
dwarf stars, neutron stars, or stellar-mass black holes would evaporate too
quickly (Maoz 1995, 1998; Genzel \etal 1997, 2000).  

      The next-best BH case is NGC 4258.  In it, a water maser disk shows
remarkably Keplerian rotation velocities inward to a radius of 
0.2 pc (Miyoshi \etal 1995).  The implied central mass, $M_\bullet = 4 \times
10^7$ $M_\odot$, again is confined to a small enough volume to exclude the above
BH alternatives (Maoz 1998).  Such arguments cannot yet be made for any other
galaxy.
Nevertheless, they increase our confidence that all of the dynamically detected
central dark objects are BHs.

      The BH search has now largely moved to HST.  With the aberrated HST, BH
work was based on indirect arguments that have serious problems (Kormendy \&
Richstone 1995).  But with COSTAR, HST has become the telescope of choice for BH
searches, and the pace of detections has accelerated remarkably. 

     The HST era is divided into two periods.  Before the installation of the
Space Telescope Imaging Spectrograph (STIS) in 1997, the main instrument used
was the Faint Object Spectrograph (FOS).  It was inefficient, because it used an
aperture instead of a slit.  Nevertheless, the first HST BH detections were made
with the FOS (M{\ts}87: Harms \etal 1994; NGC 4261: Ferrarese \etal 1996;
NGC 7052: van der Marel \& van den Bosch 1998).  It is often suggested that
HST was required to make BH cases convincing.  This is an exaggeration.  HST
beats ground-based resolution by a factor of 5, but the first BH detections made
with HST were in Virgo cluster galaxies or in ones that are 2 -- 4 times farther
away.  Virgo is $\sim$\ts20 times farther away than M{\ts}31 and M{\ts}32.
Therefore the ground-based BH discoveries in M{\ts}31 and M{\ts}32 had better 
spatial resolution (in pc) than the HST BH discoveries in the above galaxies.
Of course, the distant BHs have higher masses.  Therefore a better measure of
relative resolution is the ratio of the radius $r_{\rm cusp} = G M_\bullet /
\sigma^2$ of the BH sphere of influence to the resolution.  Table 1 lists
$r_{\rm cusp}$ for all BH detections.  Since the PSF in the ground-based
discovery observations had a
radius of $\sim 0\sd3$ -- 0\sd5 while the FOS observations used a 0\sd26
circular aperture (M{\ts}87 and NGC 7052) or a 0\sd09 square aperture (NGC
4261), Table 1 shows that the FOS BH detections in M{\ts}87, NGC 4261, and NGC
7052 had comparable or lower relative resolution than the ground-based
observations of M{\ts}31, M{\ts}32, NGC 3115, and NGC 4594.  As HST spatial
resolution improved (especially with STIS), BH cases have indeed gotten
stronger.  But the main thing that HST has provided is many more detections.

      STIS has begun a new period in the BH search.  With the efficiency of a
long-slit spectrograph and CCD detector, the search has become feasible for
most nearby galaxies that have unobscured centers and old stellar populations.
It is still not easy; finding a 10$^6$-$M_\odot$ BH is difficult at the
distance of the Virgo cluster and impossible much beyond.  But the pace of
discoveries has accelerated dramatically.  At the 2000 Summer AAS meeting, 14
new BH detections were reported, and several more have been published since.  
As a result, about 37 BH candidates are now available. We say ``about'' because
not all cases are equally strong: which ones to include is a matter of judgment.
Table 1 provides a census. 

\centerline{\null} \vfill


\includegraphics{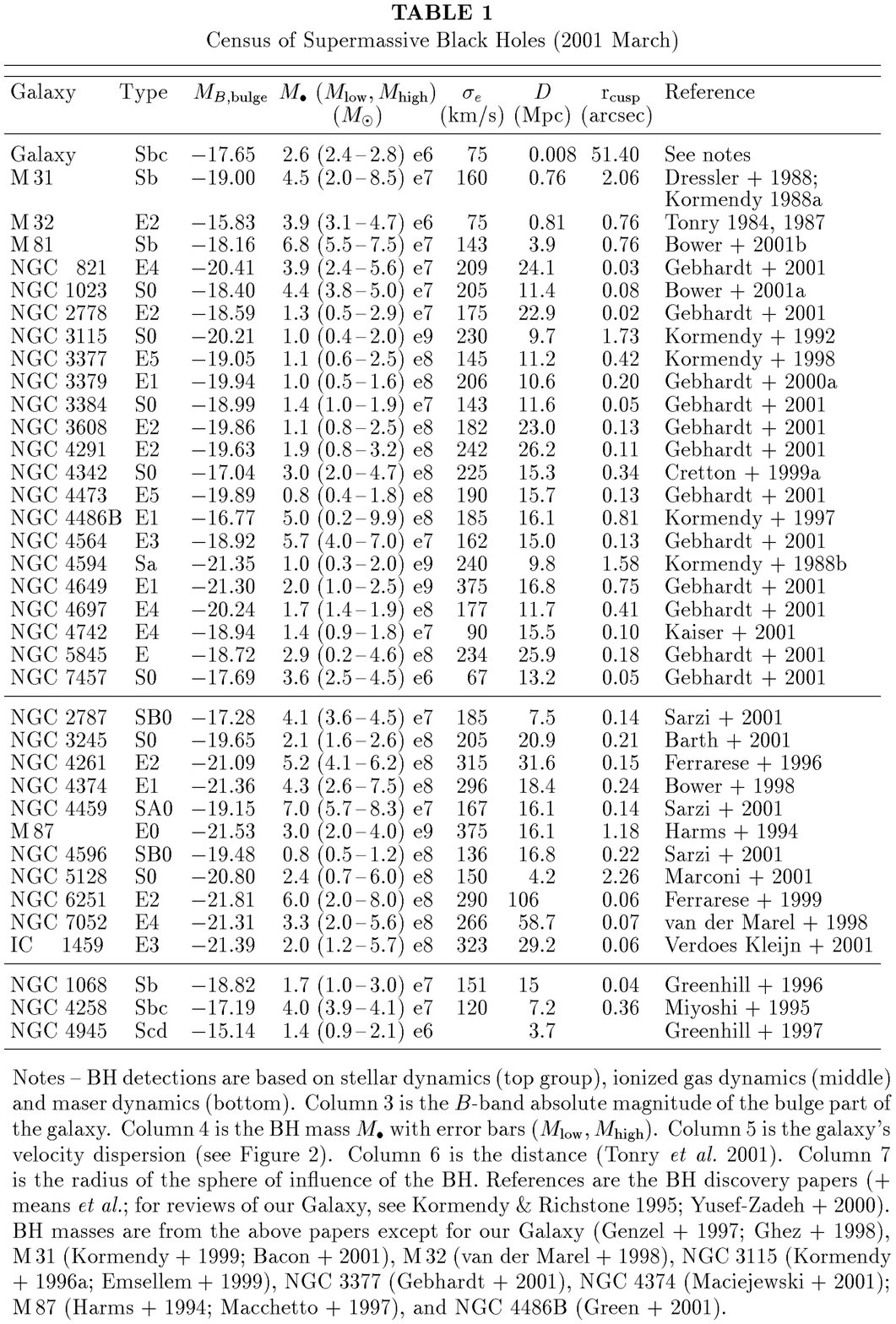}

\eject

    An important HST contribution has been to enable BH searches based on
\hbox{ionized} gas dynamics (middle part of Table 1).  The attraction of gas is
simplicity -- unlike the case of stellar dynamics, velocity dispersions are 
likely to be isotropic and projection effects are small unless the disks
are seen edge-on.  Especially important is the fact that gas disks are easy to
observe even in giant ellipticals with cuspy cores.  These galaxies are a 
problem for stellar-dynamical studies: they are expensive to observe because 
their surface brightnesses are low, and they are difficult to interpret because
they rotate so little that velocity anisotropy is very important.  It is no
accident that most BH detections in the highest-luminosity ellipticals are based
on gas dynamics.  Without these, we would know much less about the biggest BHs.

      At the same time, the uncertainties in gas dynamics are often
underestimated.  Most studies assume that disks are cold and in circular
rotation.  But the gas masses are small, and gas is easily pushed around.  It
would be no surprise to see velocities that are either slower or faster than
circular.  Faster-than-circular motions can be driven by AGN or starburst
processes, while motions that are demonstrably slower than circular are observed
in many bulges (see Kormendy \& Westpfahl 1989).  A separate issue is the
large emission-line widths seen in many galaxies.  If these are due to pressure
support, then the observed rotation velocity is less than the circular velocity
and $M_\bullet$ is underestimated if the line width is ignored.  The situation
is like that in any stellar system that has a significant velocity dispersion,
and the cure is similar.  In the context of a not-very-hot disk like that in
our Galaxy, the correction from observed to circular velocity is called the
``asymmetric drift correction'', and in the context of hotter stellar systems
like ellipticals, it is handled by three-integral dynamical models.  For gas
dynamics, the state of the art is defined by Barth \etal (2001), who discuss the
line broadening problem in detail.  They point out that asymmetric drift
corrections may be large or they may be inappropriate if the line width is due
to the internal microturbulence of gas clouds that are in individual, nearly
circular orbits.  We do not understand the physics of line broadening, so it
adds uncertainty to BH masses.  But it is likely that $M_\bullet$ will be underestimated if the line width is ignored.  In contrast, Maciejewski \& Binney (2001) emphasize that
$M_\bullet$ can be overestimated by as much as a factor of three if we neglect
the smearing effects of finite slit sizes.  The best gas-dynamic $M_\bullet$
estimates (e.{\ts}g., Sarzi \etal 2001; Barth \etal 2001) are thought to be
accurate to $\sim$\ts30\ts\%.  In future, it will be important to take all of
the above effects into account.  It is not clear {\it a priori\/} whether they
are devastating or small.  The best sign that they are manageable is the
observation that stellar- and gas-dynamical analyses imply the same $M_\bullet$
correlations (compare the squares and circles in Figure 2).

      Gas-dynamical BH searches are limited mainly by the fact that suitable gas
disks are rare.  Sarzi \etal (2001) found gas disks with well-ordered, nearly
circular velocities in only about 15\ts\% of their sample of galaxies that were
already known to have central gas.  So \lapprox \ts10\ts\% of a complete sample 
of bulges is likely to have gas disks that are usable for BH searches.   Nevertheless, within the next year, we should have gas-kinematic observations of
30\ts--\ts40 galaxies from a variety of groups.  They will provide a wealth of
information both on nuclear gas disks and on BHs.

\vfill\eject

\input cittable.tex

\section*{The $M_\bullet$ --
         $M_{B,{\rm\char'142\char'165\char'154\char'147\char'145}}$ and
         $M_\bullet$ -- $\sigma_e$ Correlations}

      The list of BHs is now long enough so that we have finished the discovery
era, when we were mainly testing the AGN paradigm, and have begun to use BH
\hbox{demographics} to address a variety of astrophysical questions.  

      Two correlations have
emerged.  Figure 2 (left) shows the correlation between BH mass and the 
luminosity of the ``bulge'' part of the host galaxy (Kormendy 1993a, Kormendy \&
Richstone 1995; Magorrian \etnuk 1998) brought up to date with new detections.
A least-squares fit gives 
\begin{equation}
M_{\bullet} = 0.78\times10^8~M_\odot\left(L_{B,\rm
bulge}\over{\hbox{$10^{10}~L_{B\odot}$}}\right)^{1.08}.
\end{equation}
Since $M/L \propto L^{0.2}$, Equation (1) implies that BH mass is proportional
to bulge mass, $M_\bullet \propto M_{\rm bulge}^{0.90}$.

      Figure 2 (right) shows the correlation between BH mass and the
luminosity-weighted velocity dispersion $\sigma_e$ within the effective radius
$r_e$ (Gebhardt {\it et nuk.} 2000b; Ferrarese \& Merritt 2000).  
A least-squares fit to the galaxies with most
reliable $M_\bullet$ measurements (Gebhardt \etnuk 2001) gives
\begin{equation}
M_{\bullet}
=1.3\times10^8~M_\odot\left(\sigma_e\over\hbox{200~{\rm km~s$^{-1}$}}\right)^{3.65}.
\end{equation}

\vfill

\includegraphics{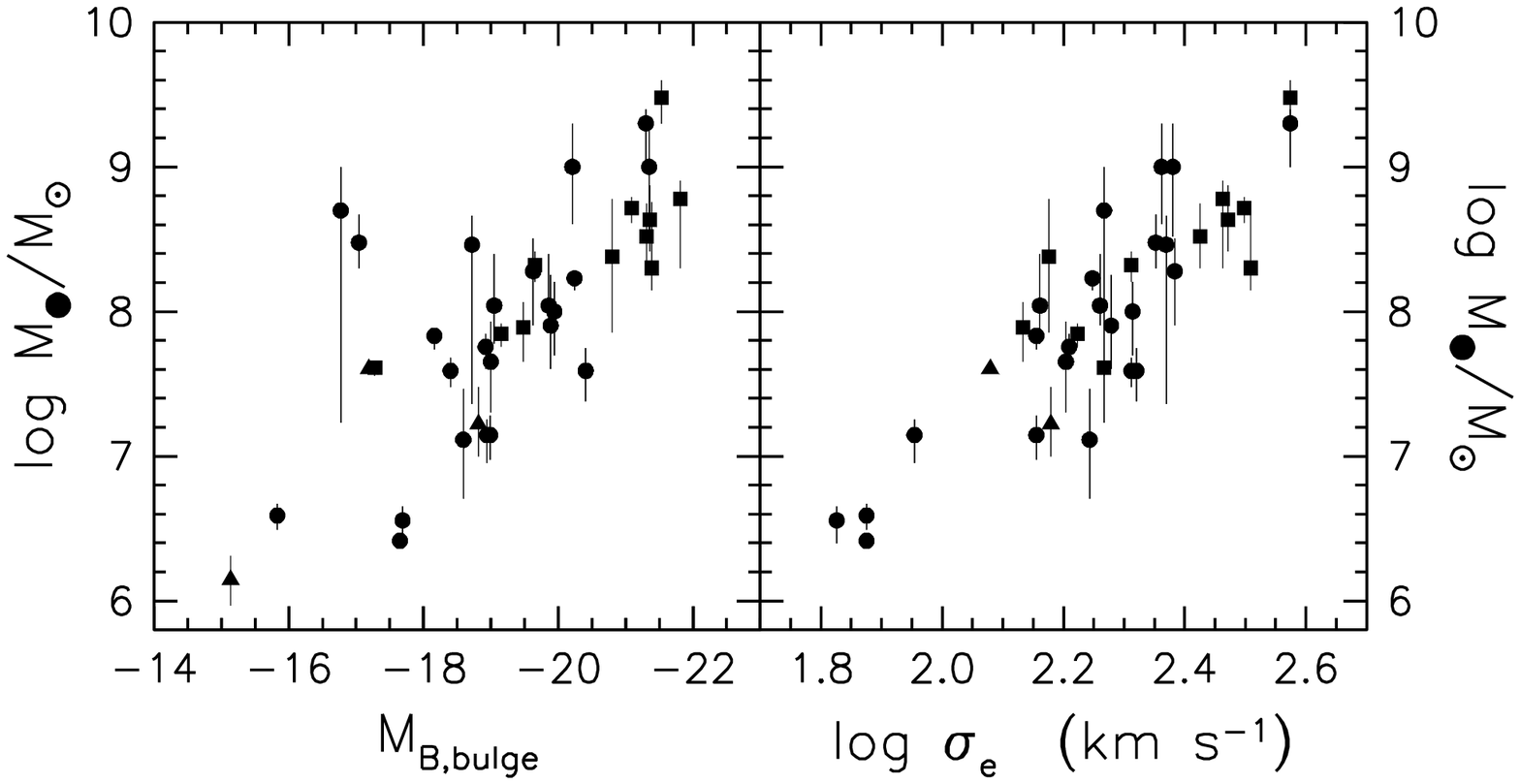}
\vskip 10pt
      {\bf FIGURE 2.} Correlation of BH mass with (left) the absolute magnitude
of the bulge component of the host galaxy and (right) the luminosity-weighted
mean velocity dispersion inside the effective radius of the bulge.  In both 
panels, filled circles indicate $M_\bullet$ measurements based on stellar
dynamics, squares are based on ionized gas dynamics, and triangles are based on
maser disk dynamics.  All three techniques are consistent with the same
correlations.  

\eject

      The scatter in the $M_\bullet$ -- $M_{B,\rm bulge}$ relation is large:
the RMS dispersion is a factor of 2.8 and the total range in $M_\bullet$ is two
orders of magnitude at a given $M_{B,\rm bulge}$.  There are also two exceptions
with unusually high BH masses.  The more extreme case, NGC 4486B (Kormendy
\etnuk 1997; Green \etal 2001), is still based on two-integral models.  Its BH
mass may decrease when three-integral models are constructed.  Despite the
scatter, the correlation is robust.  One important question has been whether the
$M_\bullet$ -- $M_{B,\rm bulge}$ correlation is real or only the upper envelope
of a distribution that extends to smaller BH masses.  The latter possibility now
seems unlikely.  Ongoing searches find BHs in essentially every bulge observed
and in most cases would have done so even if the galaxies were significantly
farther away.

      In contrast, the scatter in the $M_\bullet$ -- $\sigma_e$ correlation is
small, and the galaxies that were discrepant above are not discrepant here.
Gebhardt \etnuk (2000b) find that the scatter is consistent with the measurement
errors for the galaxies with the most reliable $M_\bullet$ measurements.  So the
$M_\bullet$ -- $\sigma_e$ correlation is more fundamental than the $M_\bullet$
-- $M_{B,\rm bulge}$ correlation.  What does this mean?  

      Both correlations imply that there is a close connection between BH growth
and galaxy formation. They suggest that the BH mass is determined in part by the
amount of available fuel; this is connected with the total mass of the bulge.  

      Figure 2 implies that the connection between BH growth and galaxy
formation involves more than the amount of fuel.  Exceptions to the
$M_\bullet$ -- $M_{B,\rm bulge}$ correlation satisfy the $M_\bullet$ --
$\sigma_e$ correlation.  This means that, when a BH is unusually high in
mass for a given luminosity, it is also high in $\sigma_e$ for that luminosity.
In other words, it is high in the Faber-Jackson
(1976) $\sigma(L)$ correlation.  One possible reason might be that the 
mass-to-light ratio of the stars is unusually high; this proves not to be
the main effect.  The main effect is illustrated in Figure 3.  Ellipticals that
have unusually high dispersions for their luminosities are unusually compact:
they have unusually high surface brightnesses and small effective radii for
their luminosities.  Similarly, cold galaxies are fluffy: they have low 
effective surface brightnesses and large effective radii for their luminosities.  
Therefore, when a galaxy is observed to be hotter than average, we conclude
that it underwent more dissipation than average and shrunk inside its dark
halo to a smaller size and higher density than average.  That is, it
``collapsed'' more than the average galaxy.   

      We can show this quantitatively by noting that the $M_\bullet$ -- 
$M_{B,\rm bulge}$ and \hbox{$M_\bullet$ -- $\sigma_e$} relations are almost
equivalent.  The left panel in Figure 2 is almost a correlation of $M_\bullet$
with the mass of the bulge, because mass-to-light ratios vary only slowly with
luminosity. Bulge mass is proportional to $\sigma_e^2 r_e / G$. So a black hole
that satisfies the $M_\bullet$ -- $\sigma_e$ correlation will look discrepant in
the $M_\bullet$ -- $\sigma_e^2 r_e$ correlation if $r_e$ is smaller or larger
than normal.  We conclude that BH mass is directly connected with the details of
how bulges form.

      This result contains information about when BHs accreted their
mass.  There are three generic possibilities.  (1) BHs could have grown to
essentially their present masses before galaxies formed and then regulated the
amount of galaxy that grew around them (e.{\thinspace}g., Silk \& Rees 1998).
(2) Seed BHs that were already present at the start of galaxy formation or that
formed early could have grown to their present masses as part of the galaxy
formation process.  (3) Most BH mass may have been accreted after galaxy
formation from ambient gas in the bulge.  The problem is that the $M_\bullet$
correlations do not directly tell us which alternative dominates.  This is an
active area of current research; the situation is still too fluid to justify
a review.  All three alternative have proponents even on the Nuker team. 

\vfill
      
\vskip 10pt

\includegraphics{collapse-bw.ps}

      {\bf FIGURE 3.}  Correlations with absolute magnitude of velocity
dispersion (upper panel), effective surface brightness (middle panel) and
effective radius (lower panel) for elliptical galaxies from the Seven Samurai
papers.  Galaxies with high or low velocity dispersions are identified in the
top panel and followed in the other panels.

\eject

      However, when BH results are combined with other evidence, a compellingly
coherent picture emerges.  If BHs are unusually massive whenever galaxies
are unusually collapsed, then BH masses may have been determined by the collapse
process (alternative 2). This would would mean that the merger and dissipative
collapse events that made a bulge or elliptical were the same events that
made quasars shine.  Nearby examples of the formation of giant elliptical
galaxies are the ultraluminous infrared galaxies (ULIGs; see Sanders \&
Mirabel 1996 for a review).  Sanders \etal (1988a, b) have suggested that ULIGs
are quasars in formation; this is essentially the picture advocated here.  Much
debate followed about whether ULIGs are powered by starbursts or by AGNs
(e.{\ts}g., Filippenko 1992; Sanders \& Mirabel 1996; Joseph 1999; Sanders
1999).  Observations now suggest that about 2/3 of the energy comes from 
starbursts and about 1/3 comes from nuclear activity (Genzel \etal 1998; Lutz 
\etal 1998).  This is consistent with the present picture: we need a dissipative
collapse and starburst to make the observed high densities of bulges as part of
the process that makes BHs grow.  Submillimeter observations are finding
high-redshift versions of ULIRGs from the quasar era (Ivison \etal 2000).
Many are AGNs.  Further evidence for a connection between ULIGs and AGN activity
is reviewed in Veilleux (2000).  ULIG properties strongly suggest that bulge
formation, BH growth, and quasar activity all happen together.  

\section*{Which Galaxies Contain BH\lowercase{s}?  Which Do Not?\\
          M$_\bullet$ Upper Limits} 

      So far, BHs have been discovered in every galaxy that contains a bulge and
that has been observed with enough resolution to find a BH consistent with the
correlations in Figure 2.  The canonical BH is about 0.13\ts\% of the mass of
the bulge; the scatter is more than a factor of 10.  Table 2 lists the strongest
BH mass limits.  We fail to find BHs in pure disk and related galaxies.  These
are discussed in the next section.

\vfill

\includegraphics{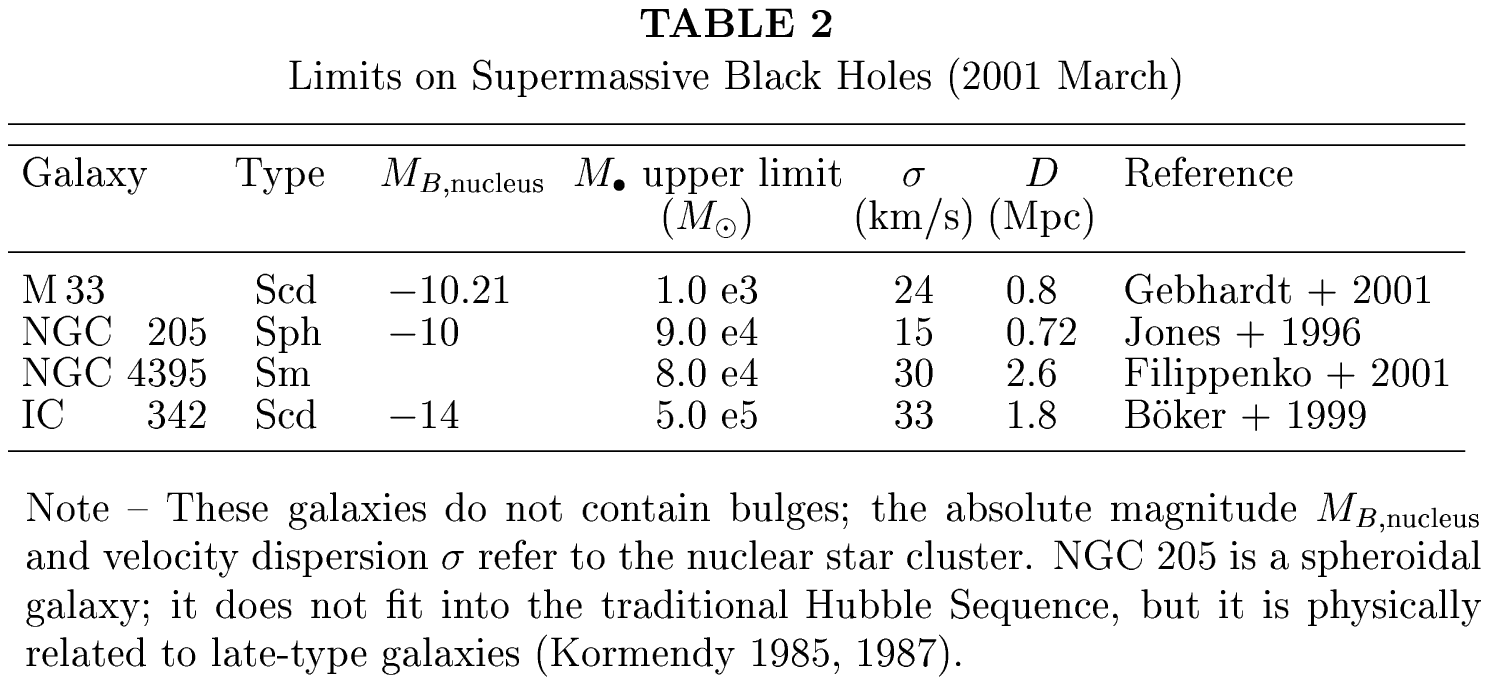}

\eject

\section*{The $M_\bullet$ -- $M_{B,{\rm total}}$ Correlation:\\BHs Do Not Know About Disks} 

\lineskip=0pt \lineskiplimit=0pt

      It is important to note that BH mass does not correlate with disks in the
same way that it does with bulges.  Figure 4 shows the correlations of BH mass
with (left) bulge and (right) total luminosity.  Figure 4 (right) shows that
disk galaxies with small bulge-to-total luminosity ratios destroy the reasonably
good correlation seen in Figure 4 (left).  In addition, Figure 4 shows
four galaxies that have strong BH mass limits but no bulges.  They further
emphasize the conclusion that disks do not contain BHs with nearly the same mass
fraction as do bulges.  In particular, in the bulgeless galaxy M{\thinspace}33,
the upper limit on a BH mass from STIS spectroscopy is $M_\bullet$
$_<\atop{^\sim}$ 1000 $M_\odot$.  If M{\thinspace}33 contained a BH with the
median mass fraction observed for bulges, then we would expect that $M_\bullet
\sim 3 \times 10^7$ $M_\odot$.

      Figure 4 tells us that BH masses do not ``know about'' galaxy disks.
Rather, they correlate with the high-density bulge-like component in galaxies.

      These results do not preclude BHs in pure disk galaxies as long as they
are small.  Filippenko \& Ho (2001) emphasize that some pure disks are Seyfert
galaxies.  They probably contain BHs.  An extreme example is NGC 4395, the
lowest-luminosity Seyfert known (Fig.~4).  However, if its BH were radiating at
the Eddington rate, then its mass would be only $M_\bullet \sim 100$ $M_\odot$
(Filippenko \& Ho 2001).  So disks can contain BHs, but their masses are {\it
much\/} smaller in relation to their disk luminosities than are bulge BHs in
relation to bulge luminosities.  It is possible that the small BHs in disks are
similar to the seed BHs that once must have existed even in protobulges before
they grew monstrous during the AGN era.

\vfill
      
\vskip 10pt

\includegraphics{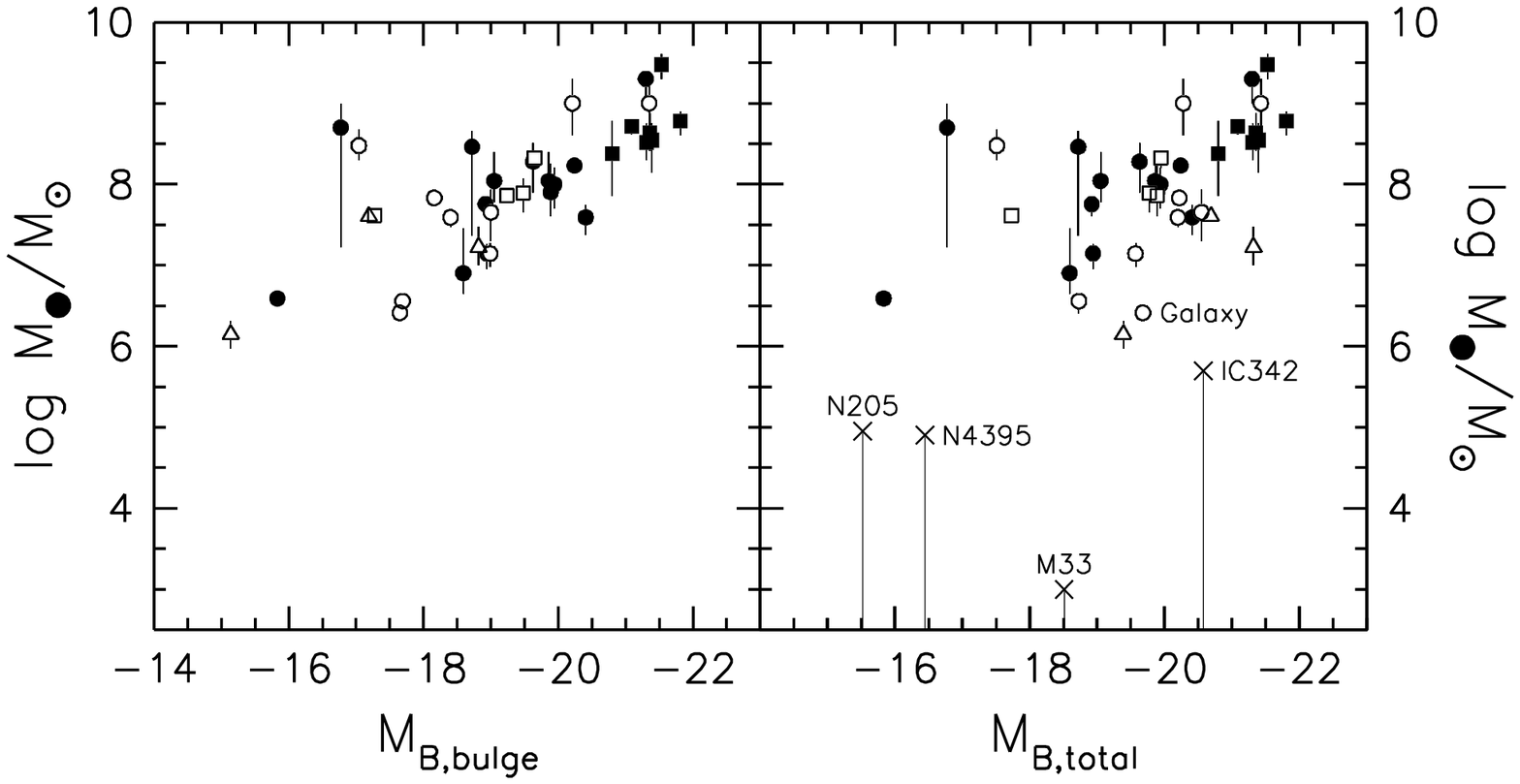}

      {\bf FIGURE 4.} (left) $M_\bullet$ -- $M_{B,\rm bulge}$ correlation from 
Figure 1.  (right) Plot of $M_\bullet$ against the total absolute magnitude of
the host galaxy.   Filled symbols denote elliptical galaxies, open symbols
denote bulges of disk galaxies.  Crosses denote galaxies that do not contain
a bulge: M{\thinspace}33 is from Gebhardt \etal (2001); IC 342 is from B\"oker
\etal (1999), and NGC 4395 is from Filippenko \& Ho (2001).

\eject

\section*{The $M_\bullet$ --
         $M_{B,{\rm\char'142\char'165\char'154\char'147\char'145}}$ 
         Correlation. II.\\Bulges Versus Pseudobulges} 

   So far, we have discussed elliptical galaxies and the bulges of disk galaxies
as if they were equivalent.  In terms of BH content, they are indistinguishable:
they are consistent with the same $M_\bullet$ -- $M_{B,\rm bulge}$ and
$M_\bullet$ -- $\sigma_e$ correlations.  But a variety of observational and
theoretical results show that there are two different kinds of high-density
central components in disk galaxies.  Both have steep surface brightness
profiles.  But, while classical bulges in (mostly) early-type galaxies are like
little ellipticals living in the middle of a disk, the ``pseudobulges'' of
(mostly) late-type galaxies are physically unrelated to ellipticals.  

      Pseudobulges are reviewed in Kormendy (1993b).  Observational evidence for
disklike dynamics includes (i) velocity dispersions $\sigma$ that are smaller
than those predicted by the Faber-Jackson (1976) $\sigma$ -- $M_B$ correlation, 
(ii) rapid rotation $V$ that implies $V/\sigma$ values above the ``oblate line''
describing rotationally flattened, isotropic spheroids in the $V/\sigma$ --
ellipticity diagram, and (iii) spiral structure that dominates the pseudobulge
part of the galaxy.  These observations and $n$-body simulations imply that
high-density central disks can form out of disk gas that is transported toward
the center by bars and oval distortions.  They heat themselves, e.{\ts}g.~by
scattering of stars off of bars (Pfenniger \& Norman 1990).
The observations imply that most early-type galaxies contain
bulges, that later-type galaxies tend to contain pseudobulges, and that only
pseudobulges are seen in Sc -- Sm galaxies.

    Andredakis \& Sanders (1994), Andredakis, Peletier, \& Balcells (1995), and 
Courteau, de Jong, \& Broeils (1996) show that the ``bulges'' of many late-type
galaxies have nearly exponential surface brightness profiles.  It is likely that
these profiles are a signature of pseudobulges, especially since blue colors
imply that they are younger than classical bulges (Balcells \& Peletier 1994).  

    HST observations strengthen the evidence for pseudobulges.  Carollo \etal
(1997, 1998a, b) find that many bulges have disk-like properties, including young 
stars, spiral structure, central bars, and exponential brightness
profiles.  It seems safe to say that no-one who saw these would suggest that 
they are mini-ellipticals living in the middle of a disk.  They look more like
late-type or irregular galaxies.  To be sure, Peletier \etal (2000) find that 
bulges of early-type galaxies generally have red colors: they are old.  True
bulges that are similar to elliptical galaxies do exist; M{\thinspace}31 and NGC
4594 contain examples.  But the lesson from the Carollo papers is that
pseudobulges are more important than we expected.  Like Kormendy (1993b) and
Courteau \etal (1996), Carollo and collaborators argue that these are not real
bulges but instead are formed via gas inflow in disks.  

    So there is growing evidence that the ``bulges'' in Fig.~2 are two
different kinds of objects.  Classical bulges are thought to form like 
ellipticals, in a dissipative collapse triggered by a merger.  Pseudobulges are
thought to form by secular evolution in disks.  In both cases, gas flows inward
and may feed BHs.  One way to explore this is to ask whether bulges and
pseudobulges have the same BH content. 

\vfill\eject 

\centerline{\null} \vfill
      
\vskip 10pt


\includegraphics{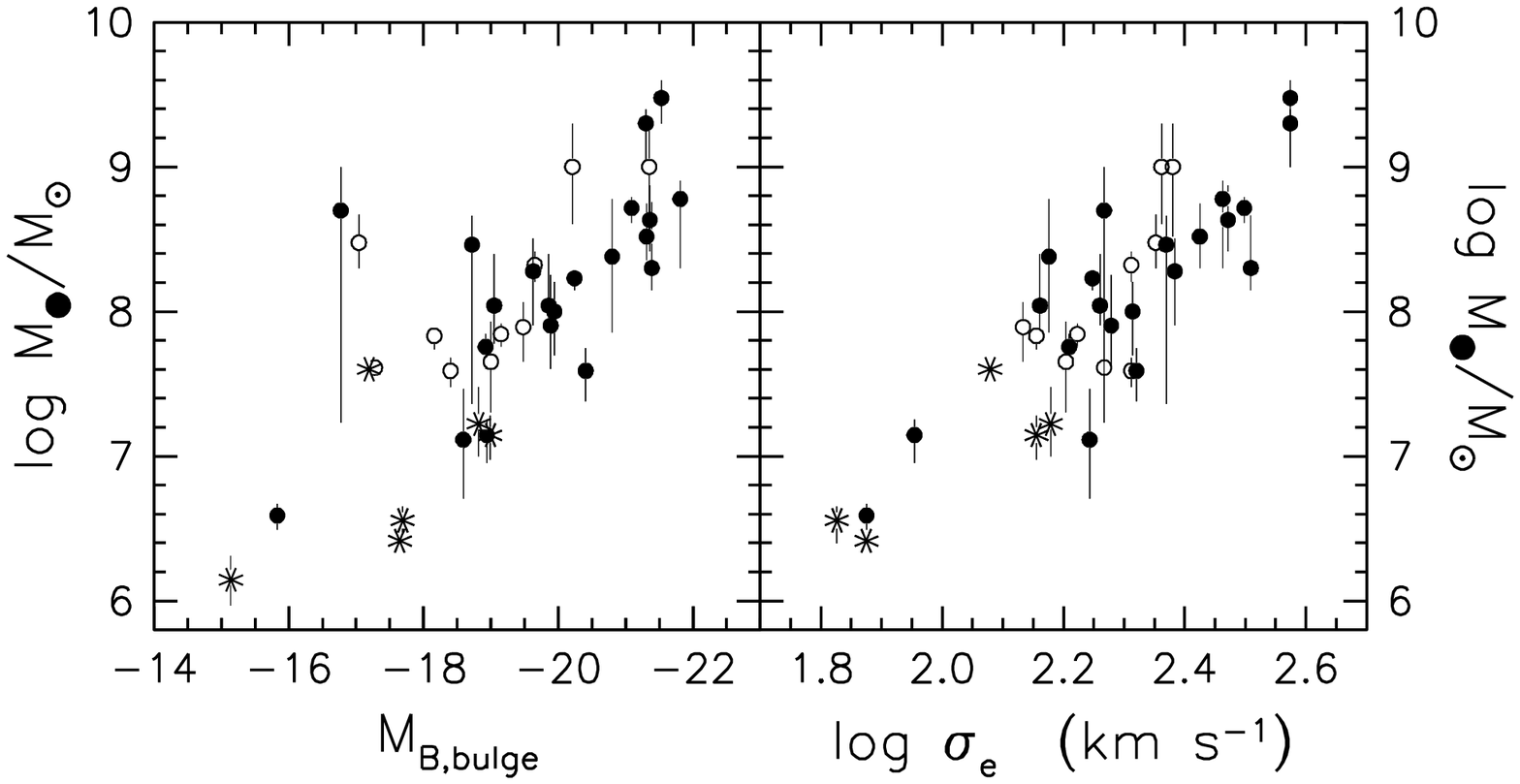}

      {\bf FIGURE 5.} The $M_\bullet$ -- $M_{B,\rm bulge}$ (left) and 
$M_\bullet$ -- $\sigma_e$ (right) correlations for elliptical galaxies (filled circles),
bulges (open circles) and pseudobulges (stars).   

\vskip 10pt

     The answer is shown in Figure 5.  Pseudobulges have relatively low
luminosities; this is plausible, since they are made from disks.  But for their
low luminosities, they have normal BH masses.  The indentification of pseudobulges 
is still somewhat uncertain, and only a few have been observed.
It will be important to check our result with a larger sample.  However, it is
consistent with the hypothesis that (pseudo)bulge formation and BH feeding are
closely connected.  Present data do not show any dependence of $M_\bullet$ on
the details of whether BH feeding happens rapidly during a collapse or slowly
via secular evolution of the disk.

      If disks contain only small BHs while the pseudobulges that form from
disks contain standard BHs with 0.13\thinspace\% of the pseudobulge mass, then
we conclude that these BHs must have grown to their present masses during
pseudobulge formation.  

      The smallest BHs provide an argument that most BH growth did not happen
after bulge formation.  Some pseudobulges are still forming now; there is little
time after bulge formation.  Also, these objects do not contain fuel in the form
of x-ray gas.  And galaxies like M{\thinspace}32 contain little gas of any sort
for late accretion.

\vskip -20pt \null

\section*{Black Holes and Galaxy Formation: Conclusion}

      Galaxy formation is complicated, so any conclusions that we reach now are
less secure than the observational results discussed above.  However: {\it The
observations suggest that the major events that form a bulge
and the major growth phases of its BH -- when it shone as an AGN -- were the
same events.}  The likely formation processes are either a series of dissipative
mergers that fuel starbursts and AGN activity (Sanders \etal 1988a, b) or
secular inward flow of gas in disks that builds pseudobulges and simultaneously 
feeds their BHs (Kormendy \etnuk 2001).

\eject

\section*{The Future}

      The future is promising: (1) the census of BHs is expected to grow 
rapidly as HST reaches its full potential and as new techniques allow us to
measure $M_\bullet$ in more distant galaxies; (2) the ongoing unification of the
BH and galaxy formation paradigms is fundamental progress, and (3) x-ray
satellites and gravitational wave detectors are expected to probe the immediate
vicinity of the Schwarzschild radius.

\subsection*{Black Holes in Distant Galaxies}

\pretolerance=10000  \tolerance=10000

      Measuring $M_\bullet$ by making dynamical models of observations that
spatially resolve the central kinematics (Tables 1 and 2) are well tested
techniques.  Confidence is growing that the resulting BH masses are accurate to
within $\sim$ 30\ts\% in the best cases.  This has allowed us to begin
demographic studies of BHs in nearby galaxies.  But the above techniques have a
fundamental limitation.  They cannot be applied unless the galaxies are close
enough so that we can spatially resolve the region that is dynamically affected
by the BH.  Within a few more years, the most interesting galaxies that are
accessible with HST resolution will have been observed, and new detections will
slow down.  Expected advances in spatial resolution will enable important but
only incremental progress.  The subject could use a breakthrough that allows us
to measure BH masses in much more distant objects.

      In this context, Figure 6 is encouraging news.  It compares BH masses
based on spatially resolved kinematics with masses derived by two other 
techniques, reverberation mapping (Blandford \& McKee 1982; Netzer \& Peterson
1997) and ionization models (Netzer 1990; Rokaki, Boisson, \& Collin-Souffrin
1992).  Both techniques have been available for some time, but it was not clear
how much they could be trusted.  Figure 6 shows that both techniques produce BH
masses that are consistent with the $M_\bullet$ correlations discussed in 
earlier sections.

      Reverberation mapping exploits the time delays measured between
brightness variations in the AGN continuum and in its broad emission lines.
These are interpreted as the light travel times between the BHs and the
clouds of line-emitting gas.  The result is an estimate of the radius $r$ of
the broad-line region.  We also have a velocity $V$ from the FWHM of the
emission lines.  Together, these measure a mass $M_\bullet \approx V^2 r / G$.
However, a number of authors (Wandel 1999; Ho 1999; Wandel, Peterson, \& Malkan
1999) have pointed out that reverberation mapping BH masses are systematically
low in the $M_\bullet$ -- $M_{B,\rm bulge}$ correlation.  Recently, Gebhardt
\etnuk (2000c; see Figure 6, below, for an update) have shown that reverberation
mapping BH masses agree with the $M_\bullet$ -- $\sigma_e$ correlation.  This
suggests that the problem uncovered in previous comparisons was that the bulge
luminosities of the reverberation mapping galaxies were measured incorrectly or
were inflated by young stars.  Gebhardt \etnuk (2000c) and Figure 6 here 
suggest that reverberation mapping does produce reliable BH masses.

\eject

\centerline{\null} \vskip 220pt
      
\vskip 9pt


\includegraphics{newion.ps}

      {\bf FIGURE 6.} The $M_\bullet$ -- $\sigma_e$ correlation for galaxies
with BH masses from detailed dynamical models applied to spatially resolved
kinematics (filled symbols as in Figure 2), reverberation mapping (crosses), and
ionization models (plus signs).   

\vskip 9pt

      Similarly, ionization model BH masses -- ones based on the observed
correlation between quasar luminosity and the radius at which the
broad-line-emitting gas lives -- are largely untested and therefore uncertain.
Laor (1998) and Gebhardt \etnuk (2001) now show that this technique also appears
to produce $M_\bullet$ values with no systematic offset from other techniques
(Figure 6).

      These results are important because neither reverberation mapping nor
ionization models require us to spatially resolve the central region affected by
the BH.  Both techniques can be applied to objects at arbitrarily large
distances.  Therefore BH masses can now be estimated for quasars out to
redshifts of nearly $z = 6$.  Ongoing surveys like 2dF and the Sloan Digital 
Sky Survey are producing thousands of quasar detections.  BH masses should
therefore be derivable for very large samples that span all redshifts from 
$z = 0$ to the most distant objects known.  It will be important to check as
well as possible that the physical circumstances that make the ionization models
work so well are still valid far away.  Nevertheless, it should be possible
to directly measure the growth of BHs in the Universe.  

\vskip -20pt \null

\subsection*{The Effects of Black Holes on Galaxy Structure}

\pretolerance=10000  \tolerance=10000

      There is other encouraging news, too.  In the past, the BH search was
decoupled from other galaxy studies.  It was carried out largely in isolation
to test the AGN paradigm.  Furthermore, the early BH detections
were mostly in inactive galaxies, so even the connection with AGN physics
was indirect.  This situation was reviewed in Kormendy \& Richstone (1995).
But now BH results are beginning to connect up with a variety of work
of galaxy physics.   The subject is large and our space is limited.  We
therefore mention briefly only three of the developing results.

      1 -- Triaxial elliptical galaxies evolve rapidly toward axisymmetry if the
central gravitational potential well gets steep enough (Lake \& Norman 1983; 
Gerhard \& Binney 1985; Norman, May, \& van Albada 1985; Valluri \& Merritt
1998; Merritt \& Quinlan 1998; Poon \& Merritt 2001; Holley-Bockelmann \etal
2001; see Merritt 1999 for a review).  This can be achieved either by increasing
the central density of stars via gas infall and star formation or by the growth
of a BH.  In either case, chaotic mixing of stellar orbits redistributes stars
in phase space and causes orbit shapes to evolve.  Box orbits, which support the
triaxial structure but which allow stars to pass arbitrarily close to the
center, are destroyed in favor of orbits that suppport axisymmetric structure.
To the extent that triaxiality promotes gas infall and BH feeding, the evolution
may also turn off the feeding when the BH has grown to 1 or 2 \% of the bulge
mass.  These processes help to explain the observed upper limit to the BH mass
fraction.

      2 -- Some elliptical galaxies have ``cuspy cores'', i.{\ts}e., density
distributions that break at small radii from steep outer power laws to shallow
inner power laws.  These cores may be produced by the orbital decay of binary
BHs (Begelman, Blandford, \& Rees 1980; Ebisuzaki, Makino, \& Okamura 1991;
Makino \& Ebisuzaki 1996; Quinlan 1996; Quinlan \& Hernquist 1997; Faber \etnuk
1997; Milosavljevi\'c \& Merritt 2001).  The formation of BH binaries is a
natural consequence of hierarchical galaxy mergers.  The orbits then decay
(i.{\ts}e., the binaries get ``harder'') by flinging stars away.  This BH
scouring may reduce the stellar density enough to produce a break in the density
profile.

      3 -- Three-integral dynamical models tell us the distribution of
stellar orbits around a BH.  Preliminary results (van der Marel \etal 1998;
Cretton \etal 1999b; Gebhardt \etnuk 2000a, 2001; Richstone \etnuk 2001)
show an important difference between core and power-law galaxies.  In core
galaxies, the central tangential velocity dispersion $\sigma_t$ is larger
than the radial dispersion $\sigma_r$.  Large tangential anisotropy is
consistent with the effects of BH binaries (Nakano \& Makino 1999a,{\ts}b)
and BH scouring (Quinlan \& Hernquist 1997).  In contrast, coreless
galaxies, which have featureless, almost power-law density profiles, are
observed to have $\sigma_r \simeq \sigma_t$. This is more consistent with
the adiabatic growth of single BHs via gas accretion (Quinlan, Hernquist,
\& Sigurdsson 1995 and references therein).  Further studies of the
relationship between BHs and properties of their host galaxies should
provide much better constraints on the relationship between BHs and galaxy
formation.

      The above developments are an important sign of the developing maturity
of this subject.  Finding convincing connections between BH properties and the
microphysics of galaxies contributes in no small measure to our confidence in
the BH picture.  The medium-term future of this subject is therefore very
promising.

      In the longer-term future, the most fundamental progress is expected to
come from gravitational wave astronomy.  We can look forward to the true 
maturity of work on supermassive BHs when the Laser Interferometer Space Antenna
(LISA) begins to provide a direct probe of strong gravity. 

\acknowledgments

      It is a pleasure to thank our Nuker collaborators for helpful discussions
and for permission to use our BH detection results before publication.  We are
also most grateful to Gary Bower, Richard Green, Mary Beth Kaiser, and Charles
Nelson for communicating STIS team BH detections before publication.

\end{document}

%% file: cittable.tex
%
%
\def\endtable{\endgroup}
\def\tableheight{\vrule width 0pt height 8.5pt depth 3.5pt}
{\catcode`|=\active \catcode`&=\active 
    \gdef\tabledelim{\catcode`|=\active \let|=\vbar
                     \catcode`&=\active \let&=\nobar} }
\def\table{\begingroup
    \def\twidth{\hsize}
    \def\tablewidth##1{\def\twidth{##1}}
    \def\defaultheight{\vrule width 0pt height 8.5pt depth 3.5pt}
    \def\heightdepth##1{\dimen0=##1
        \ifdim\dimen0>5pt 
            \divide\dimen0 by 2 \advance\dimen0 by 2.5pt
            \dimen1=\dimen0 \advance\dimen1 by -5pt
            \vrule width 0pt height \the\dimen0  depth \the\dimen1
        \else  \divide\dimen0 by 2
            \vrule width 0pt height \the\dimen0  depth \the\dimen0 \fi}
    \def\spacing##1{\def\defaultheight{\heightdepth{##1}}}
    \def\nextheight##1{\noalign{\gdef\tableheight{\heightdepth{##1}}}}
    \def\end{\cr\noalign{\gdef\tableheight{\defaultheight}}}
    \def\zerowidth##1{\omit\hidewidth ##1 \hidewidth}    
    \def\hline{\noalign{\hrule}}
    \def\skip##1{\noalign{\vskip##1}}
    \def\bskip##1{\noalign{\hbox to \twidth{\vrule height##1 depth 0pt \hfil
        \vrule height##1 depth 0pt}}}
    \def\header##1{\noalign{\hbox to \twidth{\hfil ##1 \unskip\hfil}}}
    \def\bheader##1{\noalign{\hbox to \twidth{\vrule\hfil ##1 
        \unskip\hfil\vrule}}}
    \def\spanloop{\span\omit \advance\mscount by -1}
    \def\extend##1##2{\omit
        \mscount=##1 \multiply\mscount by 2 \advance\mscount by -1
        \loop\ifnum\mscount>1 \spanloop\repeat \ \hfil ##2 \unskip\hfil}
    \def\vbar{&\vrule&}
    \def\nobar{&&}
    \def\hdash##1{ \noalign{ \relax \gdef\tableheight{\heightdepth{0pt}}
        \toks0={} \count0=1 \count1=0 \putout##1\end 
        \toks0=\expandafter{\the\toks0 &\end} \xdef\piggy{\the\toks0} }
        \piggy}
    \let\e=\expandafter
    \def\putspace{\ifnum\count0>1 \advance\count0 by -1
        \toks0=\e\e\e{\the\e\toks0\e&\e\multispan\e{\the\count0}\hfill} 
        \fi \count0=0 }
    \def\putrule{\ifnum\count1>0 \advance\count1 by 1
        \toks0=\e\e\e{\the\e\toks0\e&\e\multispan\e{\the\count1}\leaders\hrule\hfill}
        \fi \count1=0 }
    \def\putout##1{\ifx##1\end \putspace \putrule \let\next=\relax 
        \else \let\next=\putout
            \ifx##1- \advance\count1 by 2 \putspace
            \else    \advance\count0 by 2 \putrule \fi \fi \next}   }
\def\tablespec#1{
    \def\vdimens{\noexpand\tableheight}
    \def\tabby{\tabskip=0pt plus100pt minus100pt}
    \def\r{&################\tabby&\hfil################\unskip}
    \def\c{&################\tabby&\hfil################\unskip\hfil}
    \def\l{&################\tabby&################\unskip\hfil}
    \edef\templ{\noexpand\vdimens ########\unskip  #1 
         \unskip&########\tabskip=0pt&########\cr}
    \tabledelim
    \edef\body##1{ \vbox{
        \tabskip=0pt \offinterlineskip
        \halign to \twidth {\templ ##1}}} }